\documentclass[journal,comsoc]{IEEEtran}
\usepackage{graphicx}
\usepackage{epstopdf}
\usepackage{graphicx}
\usepackage{amsmath}
\usepackage{amsfonts}
\usepackage{amssymb, color, bm}

\usepackage{algorithmicx}
\usepackage[noend]{algpseudocode}
\usepackage[T1]{fontenc}

\usepackage{multicol}
\usepackage{lipsum}
\usepackage{mwe}

\usepackage[utf8]{inputenc}
\newcommand{\norm}[1]{\left\lVert#1\right\rVert}

\newcommand{\AK}[1]{\textcolor{cyan}{#1}}

\allowdisplaybreaks
\usepackage{cite}

\usepackage{algorithm}
\usepackage[noend]{algpseudocode}
\usepackage[
  colorinlistoftodos, 
  textwidth=\marginparwidth, 
  textsize=scriptsize 
  ]{todonotes}

\title{Inference over Wireless IoT Links with Importance-Filtered Updates}

\author{Ivana Nikoloska,~\IEEEmembership{Graduate Student Member,~IEEE,}
        Josefine Holm,~\IEEEmembership{Student Member,~IEEE,}
        Anders E. Kal\o r,~\IEEEmembership{Student Member,~IEEE,}
        Petar Popovski,~\IEEEmembership{Fellow,~IEEE,}
        and~Nikola Zlatanov,~\IEEEmembership{Member,~IEEE.}
\thanks{I. Nikoloska and N. Zlatanov are with the Department
of Electrical and Computer Systems Engineering, Monash University, Melbourne,
VIC, Australia e-mails: \{ivana.nikoloska, nikola.zlatanov\}@monash.edu.}
\thanks{J. Holm, A. E. Kal\o r and P. Popovski are with the Department of Electronic Systems, Aalborg University, Aalborg, Denmark e-mails: \{jho, aek, petarp\}@es.aau.dk.}
\thanks{The work by J. Holm, A. E. Kal\o r and P. Popovski has been supported by the Danish Council for Independent Research, Grant Nr. 8022-00284B SEMIOTIC.}%
}

\begin{document}

\maketitle

\begin{abstract}
We consider a communication cell comprised of Internet-of-Things (IoT) nodes transmitting to a common Access Point (AP). The nodes in the cell are assumed to generate data samples periodically, which are to be transmitted to the AP. The AP hosts a machine learning model, such as a neural network, which is trained 
on the received data samples to make accurate inferences. We address the following tradeoff: The more often the IoT nodes transmit, the higher the accuracy of the inference made by the AP, but also the higher the energy expenditure at the IoT nodes.
We propose a data filtering scheme employed by the IoT nodes, which we refer to as distributed importance filtering in order to filter out redundant data samples already at the IoT nodes. The IoT nodes do not have large on-device machine learning models and the data filtering scheme operates under periodic instructions from the model placed at the AP. The proposed scheme is evaluated using neural networks on a benchmark machine vision dataset, as well as in two practical scenarios: leakage detection in water distribution networks and air-pollution detection in urban areas. The results show that the proposed scheme offers significant benefits in terms of network longevity as it preserves the devices' resources, whilst maintaining high inference accuracy. Our approach reduces the the computational complexity for training the model and obviates the need for data pre-processing, which makes it highly applicable in practical IoT scenarios.
\end{abstract}
\begin{IEEEkeywords}
Distributed Importance Filtering, Internet of Things, Machine Learning, Rate Reduction.
\end{IEEEkeywords}

\section{Introduction}
Machine Learning (ML) is considered to be one of the biggest innovations since the microchip and it is well on its way to becoming one of the most prolific technologies of this century. 
From image recognition and natural language processing to intelligent business and production processes, ML and deep learning in particular are transforming almost every aspect of our daily lives \cite{acc2019}. The progress we have witnessed during this last decade has been fuelled mainly by the convergence of three crucial factors, without any of which this progress would have been nearly impossible. These factors are the availability of computing power, algorithmic improvements, and data, where the latter is often considered to be today's most valuable commodity. 
A major pipeline for acquiring data is Internet of Things (IoT) \cite{gupta2016abc}. In fact, the IoT and deep learning have a two-way symbiotic relationship. On one hand, the IoT is one of the main benefactors of deep learning, as it constantly produces a vast amount of data. On the other hand, the IoT ecosystem may benefit significantly from the fact that deep learning can turn the data into insights and actions for improving the IoT-powered processes and services. 

In the IoT, a growing army of sensors capable of registering locations, voices, faces, audio, temperature, sentiment, health and the like operate autonomously with little to no human intervention. IoT nodes are typically heavily constrained in terms of hardware, computational abilities, and energy supply. In particular, IoT nodes are battery powered and for many IoT applications these simple nodes are expected to operate for at least 10 years without battery recharging or battery replacement \cite{gubbi2013internet}. In spite of these significant challenges, the number of connected IoT devices is continuing to grow and consequently so does the amount of available data from IoT nodes used for training ML models.
However, more training data does not necessarily mean better data, but it certainly means a higher computational cost. In addition, using large training datasets has many ripple effects, from an increased memory cost to a non-negligible environmental impact \cite{strubell2019energy}.
Luckily, when working with a ML model, many training data samples are redundant and thus can be ignored without impacting the final model and the inference ability of the algorithm \cite{schroff2015facenet}. This is due to the fact that many training data samples are either not informative or can already be properly handled by the ML model.

\begin{figure}[tbp]
\centering
\includegraphics[width=0.8\linewidth]{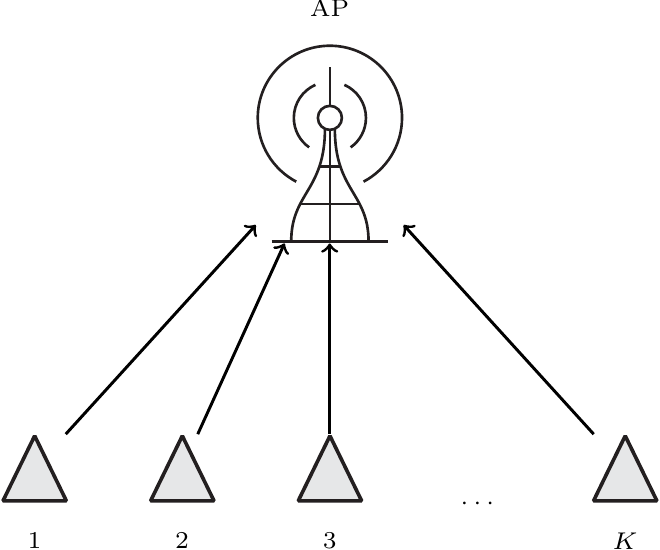} 
\caption{System model comprised of an AP and $K$ IoT nodes.}
\label{sysmodel}
\end{figure}

Motivated by this idea, in this paper, we investigate how to reduce the size of large training datasets in order to produce a smaller dataset by the IoT nodes without impacting the inference accuracy of the ML model. To this end, we consider a system model comprised of IoT nodes and an Access Point (AP), as illustrated in Fig.~\ref{sysmodel}. All IoT nodes are assumed to be typical in the sense that they face severe constraints on computation, energy consumption, and memory. 
Contrary to the IoT nodes, the AP is assumed to have abundant resources in terms of computation and memory. The IoT devices are assumed to monitor their environment and generate data samples that are to be transmitted to the AP, where they are labeled and used to train a ML model. The more training data samples are transmitted from the IoT nodes to the AP, the more accurate the model at the AP becomes  at the cost of a higher packet rate, which leads to energy expenditure at the IoT nodes   thereby reducing the lifetime of the IoT devices. Hence, there exists a trade-off between, on the one hand, the inference accuracy of the neural network at the AP  and, on the other hand, the lifetime of the IoT devices. In order to optimize this trade-off, we propose \emph{distributed importance filtering} of the data generated by the IoT nodes. The basic idea is to filter out training data samples at the IoT nodes which would most likely be redundant for training the model at the AP, such that these training data samples are not sent to the AP, thereby preserving the energy of the IoT nodes. In addition, each training sample needs to be labeled to provide context so that the model can learn from it. For example, labels might indicate whether a photo contains a bird or car, which words were uttered in an audio recording, or if an X-ray contains a tumor. Whilst data labeling is necessary for a variety of use cases including computer vision, natural language processing, and speech recognition, it is very costly. Thereby, filtering out some of the training samples also reduces the cost associated to labeling.

A useful tool which allows for filtering data samples is coreset construction \cite{feldman2011unified}. 
Coresets are succinct, small summaries of large datasets, created in a way to ensure that the optimal solution (i.e., the model parameters) found based on the summary is provably competitive with the solution found on the full dataset. Coresets were originally studied in the context of computational geometry and older approaches often relied on computationally expensive methods such as exponential grids.
We are interested in constructing coresets based on importance sampling \cite{schroff2015facenet}, \cite{loshchilov2015online}, where we neither have the full dataset in advance, nor do we make any assumptions on the probability distribution of the data. Thereby, we construct our coreset in a streaming setting, where the decision on whether a training  data samples should be transmitted to the AP is made in real time without any a priori knowledge of the dataset.

Previous works have also considered using important examples to speed up the training process of the model, e.g., \cite{katharopoulos2018not}. However, the scheme in
\cite{katharopoulos2018not} requires the full training dataset to be a priori available, so that the model can be pre-trained on multiple batches comprised of uniformly chosen samples, in order to obtain an estimate on the sampling distribution. 
In this paper, we attempt to preserve the resources of the IoT nodes, and as a result the full dataset is not accessible a priori, and we sample the data in an online manner which  ensures low energy cost.
In addition, unlike edge intelligence paradigms in which the edge nodes host neural networks, such as in \cite{mcmahan2016communication}, \cite{konevcny2016federated},\cite{bhardwaj2019memory}, we consider generic IoT devices without any intelligence, with limited computational abilities, and limited energy supply. By employing the proposed filtering scheme, the longevity of the IoT nodes is drastically improved  compared to the case when all samples are transmitted. In addition, the computational complexity for training the model is reduced, compared to the case when all samples are used. Thereby, additional data pre-processing steps aiming to decrease the training dataset size are no longer needed, all at the expense of only a small reduction in inference accuracy at the AP. We evaluate the proposed scheme on real-life problems, namely leakage detection in water distribution networks and air-pollution detection in urban areas, as well as on the machine vision dataset MNIST \cite{mnist}. Results confirm that our scheme offers significant benefits in terms of network longevity, whilst maintaining high inference accuracy. 

The rest of the paper is organised as follows. The system model is presented in Section~\ref{sys_mod}, and the considered problem is formalised in Section~\ref{prob_form}. The proposed algorithm is presented in detail in Section~\ref{prop_algo}, and it is evaluated in Section~\ref{num_eval}. A short discussion concludes the paper is Section~\ref{discus}.

\section{System Model}\label{sys_mod}
We consider a system model comprised of $K$ IoT nodes transmitting to a common AP, as shown in Fig.~\ref{sysmodel}. The IoT nodes are assumed to generate independent and identically distributed data samples denoted by $\mathcal{D}_{\text{IoT}}=\{x_1,x_2,\ldots\}$, where each sample is drawn from the distribution $p(x)$. Associated with each data sample $x_t$ is a corresponding label, $y_t\sim p(y|x)$, which is unobservable at the IoT node but assigned by the AP (without error, e.g. by a domain expert). We denote the dataset including the labels by $\mathcal{D}=\{(x_1,y_1),(x_2,y_2)\ldots,\}$, where $(x_t,y_t)\sim p(x,y)=p(x)p(y|x)$.
Due to the high cost of labeling observations, the aim is to learn a function $\psi_{\bm{\omega}}(x)$ parameterized by $\bm{\omega}$, that predicts the target value for a given input, $x$, by minimizing the generalization error
\begin{align} \label{opt}
\underset{\bm{\omega}}{\text{min}}\ P_{\epsilon}=E_{(X,Y)\sim p(x,y)}\left[\mathcal{L} \left(Y, \psi_{\bm{\omega}}(X)\right)\mid\mathcal{D}\right],
\end{align}
where $\mathcal{L}$ is a loss function. However, we assume that only a fraction of the data samples generated by the IoT nodes are sent to the AP, whereas the remaining data samples are filtered out by the IoT nodes. As a result, the generalization error must be minimized using only a subset of the available data samples $\mathcal{D}$. More specifically, during a time interval of length $T$, an IoT node is assumed to generate $m$ data samples and transmit to the AP $b$ data samples, such that $b \leq m$. We assume that the transmitted data samples are placed as payloads in data packets which are then transmitted to the AP. The packet rate $R$ is thus given by $R = b / T$, expressed in number of data samples per time unit. We further assume that the AP can provide feedback to the IoT node. This feedback is then used by the IoT nodes in order to determine which data samples should be transmitted to the AP. The data packets, both uplink and downlink, are assumed to be received and decoded without errors.

We denote the data samples received at the AP from the IoT nodes (including the labels assigned by the AP) by $\tilde{\mathcal{D}}=\{(\tilde{x}_1,\tilde{y}_1),(\tilde{x}_2,\tilde{y}_2),\ldots\}$, and assume that the samples are aggregated in batches $\bar{x}, \bar{y}$. To learn the approximation $\psi_{\bm{\omega}} (x)$ that minimizes the loss function $\mathcal{L}$, the weights of the model $\bm{\omega}$ are iteratively adjusted when the AP is presented with a new sample batch using the Stochastic Gradient Descend (SGD) algorithm  according to
\begin{align} \label{update_rule_weights}
    \bm{\omega} \leftarrow \bm{\omega} - \eta \nabla_{\bm{\omega}} \mathcal{L} \left(\bar{y}, \psi_{\bm{\omega}}(\bar{x}) \right),
\end{align}
where $\eta$ denotes the learning rate.

\section{Problem Formulation}\label{prob_form}
\begin{figure}[tbp]
\centering
\includegraphics[width=2.9in,height=6cm]{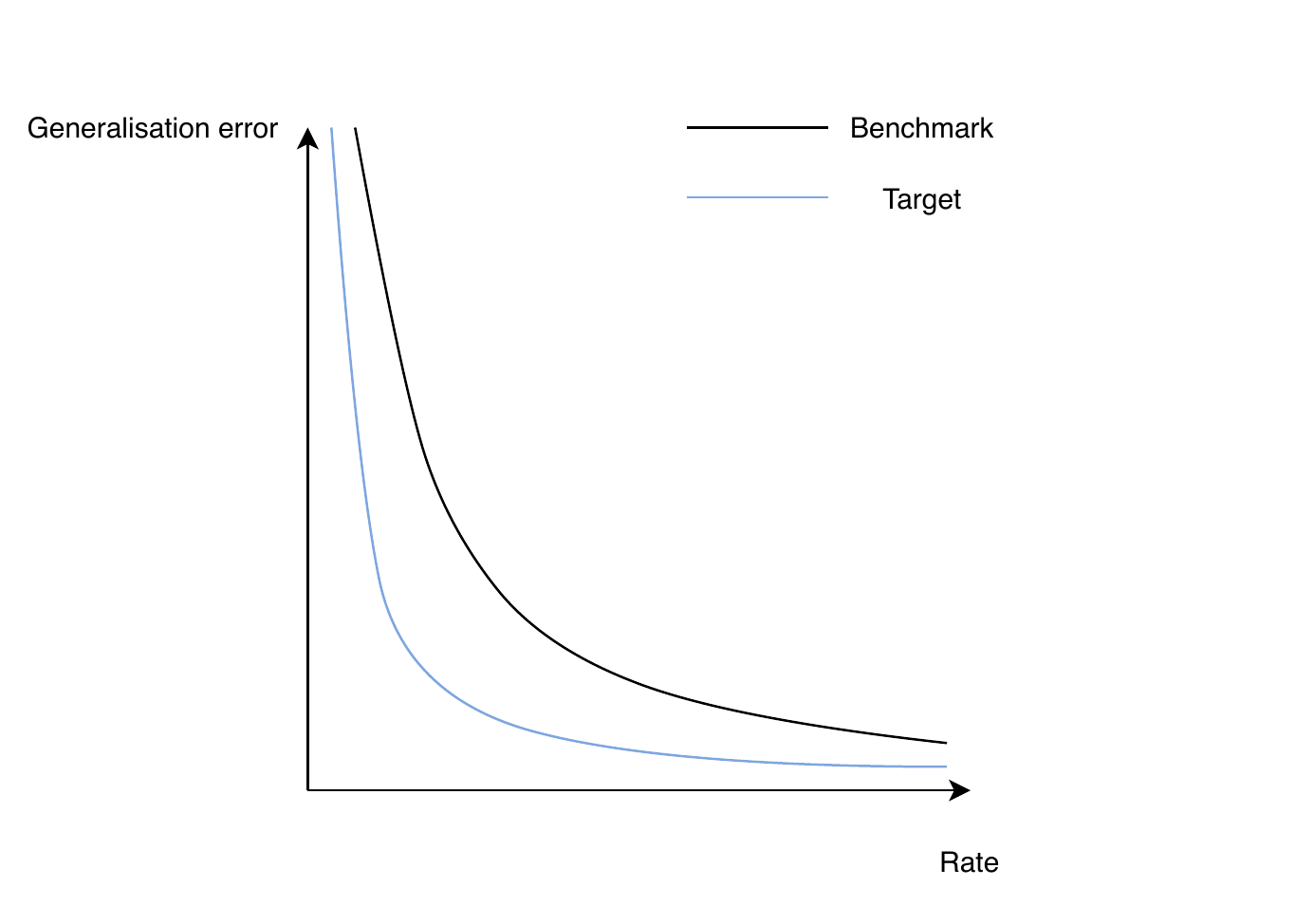}
\caption{Dependency between the generalization error $P_{\epsilon}$ and the rate $R$, for a fixed training time $t$.}
\label{num1}
\end{figure}

The generalization error $P_{\epsilon}$ is approximately inversely proportional to the logarithm of the number of training data samples~\cite{hestness2017deep}. 
As a result, for a fixed training time $T$, $P_{\epsilon}$ is approximately asymptotically related to the packet rate $R$ as
\begin{align}\label{eq_1}
    P_{\epsilon} \sim \alpha R^\beta.
\end{align}
Thus, there exists a trade-off between the training accuracy and the network longevity  since 
higher rates $R$ lead to lower generalization error, see Fig.~\ref{num1}; however, a higher $R$ also leads to a decrease of the network longevity. 
The resulting trade-off motivates us to address the role of the rate $R$ on the accuracy of the learned model and the longevity of the IoT nodes. 
To this end, we seek a filtering scheme
\begin{equation}
    \pi(x)=\begin{cases}
    1 & \text{if $x$ should be transmitted,}\\
    0 & \text{otherwise,}
    \end{cases}
\end{equation}
that carefully selects which data samples should be transmitted to the AP for training and which should be filtered out. Note that $\pi(x)$ does not depend on the labels $y$, which are assumed to be assigned by the AP and thus not available at the IoT node.
Specifically, we aim to solve the following optimization problem
\begin{align}\label{prob}
   & \underset{\pi}{\text{min}} \,\,\,\, P_{\epsilon} \nonumber \\
   & \text{s.t.} \,\,\,\, E[\pi(x)] \le R. 
\end{align}

\section{Proposed Algorithm}\label{prop_algo}
Motivated by the fact that in a large-scale setting, much of the training data is often redundant, and only a small subset of the training data is responsible for improving the training accuracy of the model \cite{schroff2015facenet}, we propose an algorithm that aims to filter redundant samples at the IoT nodes.
The method, which can be thought of as a pre-processing step in classical data-science, constructs a  coreset, i.e.,  a small,   subset of the training data that approximates the full training dataset, which is a method that can be used in many standard inference procedures to provide posterior approximations with guaranteed quality.
\subsection{Proposed Algorithm}\label{algo}
Consider the weights update algorithm of the model given by \eqref{update_rule_weights}. Intuitively, this algorithm  works as follows. In each SGD iteration, the changes in the   weights $\bm{\omega}$ depended on an "error signal",  given by $\nabla_{\bm{\omega}} \mathcal{L} \left(y, \psi_{\bm{\omega}}(x)\right)$. Training samples  that are already correctly handled by the model will not induce changes to the weights $\bm{\omega}$. As a result, feeding more of such training samples into the model  brings little to no benefit to the training accuracy of the model and only decreases the longevity of the IoT nodes. On the other hand, there are training  samples that when presented to the model cause  
large changes to the weights $\bm{\omega}$ and, as a result,  these training samples   are very informative for and by the model. 

Our aim is to develop an algorithm that will enable the IoT nodes to detect the informative training   samples and send them to the AP, while prevent wasteful transmissions of non-informative training samples.
To this end,  we introduce the \emph{leverage score function} of a sample $(x,y)$, denoted by $S(x,y)$, defined as
\begin{align} \label{lev_scores}
   S(x,y) = \norm{ \nabla_{\bm{\omega}} \mathcal{L} \left(y, \psi_{\bm{\omega}}(x) \right) }_2,
\end{align} 
where $\norm{\cdot}_2$ is the 2-norm. 
The leverage scores are computed by the AP, and broadcasted to the IoT nodes after each time interval of length $T$. We assume that each IoT node stores the $P$ most recent leverage  scores and their respective data points, $\{(\hat x_1,S(\hat x_1, \hat y_1)), \ldots, (\hat x_P,S(\hat x_P, \hat y_P))\}$, and uses them to estimate the  leverage   scores of new arriving training samples. Without loss of generality, at $t = 0$, the initial $P$ leverage scores are set to \AK{1}. 
Let $x_t$ denote the data sample arriving at a given IoT node at time $t$, and let $\hat x_p$ denote the $p$-th stored data sample with corresponding leverage score $S(\hat x_p,\hat y_p)$, for $p=1,2,..,P$. Next, let   $d ({x_t}, \hat x_p)$ denote the Euclidean distance between $x_t$  and $\hat x_p$\footnote{Note that, the distance metric can be adapted according to the application being considered, e.g., Hamming distance can be used for text classification etc.}, given by 
\begin{align}\label{Edist}
    d ({x_t}, \hat x_p) = \norm{x_t - \hat x_p}_2 
\end{align}
where $n$ is the size of each sample. Now, without loss of generality, let $\hat x_1$, $\hat x_2$, ...,$\hat x_L$ denote  the $L$  nearest neighbours of $x_t$  with respect to \eqref{Edist}, where $L<P$. Then, we define 
the estimated leverage score of the training sample $x_t$  as
\begin{align}\label{kLS}
    \hat S(x_t) = \frac{1}{L} \sum_{l = 1}^L S(\hat x_l,\hat y_l).
\end{align}
Note that this estimate, contrary to the exact leverage score, is computed without the class label $y_t$ and the model $\psi_{\omega}(x_t)$, which are not available at the IoT node. Next, the IoT node decides if  $x_t$ should be transmitted to the AP or not based on the following probability
\begin{equation} \label{q}
    q(x_t)=\frac{\exp \left(\beta(t)  \hat S(x_t) \right)}{\sum_{i=1}^P \exp \left(\beta(t)  \hat S(\hat x_i)\right)}.
\end{equation}
Note that other functions could be used in place of the exponential function to control how the leverage scores are weighted. However, the exponential function has proved to work well in practice.
In \eqref{q}, $\beta(t)$ is a factor that determines how much prioritisation is used. In particular, we define a time schedule on $\beta(t)$ so that $\beta(t)$ increases from zero to one (or close to one) over time. The rationale behind this is the fact that during the first few time intervals of length $T$, the model will produce large gradients, due to the fact that it has not been stabilised (and not necessarily because the samples are important). Thereby, by setting $\beta(t=0)$ to zero, the initial $b$ samples are transmitted with uniform probability.
Then, $\beta$ is increased after each time interval of length $T$ as
\begin{align}
    \beta (t) = \beta_{max} - \left(\beta_{max} - \beta_{min}\right)\left(\frac{\mathcal{T}-\left\lceil\frac{t}{T}\right\rceil}{\mathcal{T}}, 0\right)^+,
\end{align}
where $\mathcal{T}$ denotes the number of time intervals of length $T$ after which $\beta (t)$ reaches $\beta_{max}$. Thereby, $\beta(t)$ remains constant during each time interval of length $T$, and it is increased for the following interval. As $\beta (t)$ is increased, we anneal the amount of uniform transmissions, and transition to pure, greedy, importance-based transmissions in a slow and controlled fashion.

Thereby, each node transmits new $b \ll m$ data samples to the AP during each time interval of length $T$. Upon reception, the model weights are updated, and the leverage scores for the received samples are updated according to (\ref{lev_scores}). Then, $P \ll Kb$ updated leverage scores are transmitted back to the nodes, and this procedure is repeated until convergence. 

\subsection{Bounds on the Estimation Error}\label{risk_bounds}
The convergence rates for non-parametric regression such as \eqref{kLS} are well known and understood in the asymptotic regimes. Indeed, when $P$ is large (e.g., $P \sim Kb$), the estimate of the leverage score $\hat S (x_t)$ equals the true leverage score (that would have been obtained by the model) of sample $(x_t,y_t)$, $S (x_t, y_t)$, with probability close to one \cite{cover1967nearest}. However, we are interested in the non-asymptotic regimes since due to the previously discussed memory constraints, the nodes in the cell are often not able to store all leverage scores, and as a result the estimation precision can be affected. To bound the estimation error, let $\mathcal{B} (x_t, r)$ denote a closed-ball of radius $r$ defined as
\begin{align}
    \mathcal{B} (x_t, r) = \left\{(x,y) \in \mathcal{D} \,\,\, : \,\,\, |x_t - x| \le r \right\}.
\end{align}
Let $u_f (x_t, y_t, r)$ denote the modulus of continuity, where the following holds
\begin{align}\label{hatU}
    \hat{u}_f (x_t, y_t, r) = \sup_{(x,y) \in \mathcal{B} (x_t, r)} \left( S (x,y) - S (x_t,y) \right)
\end{align}
and
\begin{align}\label{checkU}
    \check{u}_f (x_t, y_t, r) = \sup_{(x,y)\in \mathcal{B} (x_t, r)} \left( S (x_t,y_t) - S (x,y) \right)
\end{align}
In addition, let $\eta_b$ and $\eta_v$ denote the bias and variance terms, defined as 
\begin{align}\label{bias}
    \eta_b = \left(\frac{L}{P}\right)^{1/n},
\end{align}
and
\begin{align}\label{variance}
    \eta_v = \sqrt{\frac{\log P}{L}},
\end{align}
respectively. Then, in a similar fashion to \cite{jiang2019non}, we can prove that with probability at least $1-\delta$, $\delta \to 0$ the following holds uniformly for any $(x,y) \in \mathcal{D}$
\begin{align}\label{bounds}
    & \hat S (x_t) \leq S (x_t,y_t) + \hat{u}_f (x_t,y_t \eta_b) + \eta_v \nonumber\\
    & \hat S (x_t) \geq S (x_t,y_t) - \check{u}_f (x_t,y_t \eta_b) - \eta_v,
\end{align}
when $L \geq \log(p)$ holds. In \eqref{bounds}, $\hat{u}_f (x_t,y_t,\eta_b)$, $\check{u}_f (x_t,y_t \eta_b)$, $\eta_b$ and $\eta_v$ can be found from \eqref{hatU}, \eqref{checkU}, \eqref{bias}, and \eqref{variance}, respectively.

Next, let $r_L (x_t)$ denote the radius within which the $L$ nearest neighbours of $x_t$ lie (including $x_t$). Following Lemmas 7 and 8 in \cite{chaudhuri2010rates}, we can bound $r_L (x_t)$ by $r = (L/P)^{1/n}$, meaning, every point $x_t$ will have at least $L$ neighbours within radius $r$ with probability close to one. 
This bound on $r_L (x_t)$ can be used to tighten the estimates of the leverage scores. In particular, we cluster the stored samples, which ensures that $ r_L (x_t) \ll \left(L/P\right)^{1/n}$ and thereby the neighbours will never be as far as $\left(L/P\right)^{1/n}$. To do so, let $\bar{x}$ denote a super-sample comprised by the AP from the previously transmitted $b$ samples from node $k$ as
\begin{align}
    \bar{x} = \frac{1}{b} \sum_{i = 1}^b x_i.
\end{align}
The AP calculates the super-samples $\bar{x}, \forall k$ and also computes the distances akin to \eqref{Edist} for the super-samples. The AP then transmits the leverage scores to node $k$ corresponding to the samples which comprise the $\left\lceil \frac{P}{b} \right\rceil$ nearest neighbours of super-sample $\bar{x}$. 
As a result, the obtained leverage scores via \eqref{kLS} are good estimates of the exact leverage scores obtained by the model itself and the important samples can be extracted even in cases when there is little structure in the data.

Note that, by employing the proposed scheme, additional data pre-processing steps aimed at reducing the dataset size are no longer needed, as the pre-processing is essentially carried out by the IoT nodes themselves. 
Further, since fewer packets are transmitted, the energy consumption profile of distributed importance filtering is much better than classical schemes where all packets are transmitted. This in practice translates to consuming energy in each time slot $t$ only for waking up, and for reception every $T$ time slots. The former is due to the fact that the nodes need to make a measurement and decide whether to transmit or not. If not, the node immediately initiates a turn off sequence and goes back to sleep, without engaging the radio. Thereby, the node does not consume energy for radio preparation, transmission, acknowledgement reception, and turning off the radio. The later is due to the fact that the IoT nodes need to receive the leverage scores, for which the radio is engaged. As these updates occur every $T$ time slots and as a result are far less frequent than potential transmissions, the energy consumption of the proposed scheme is lower compared to the case when all packets are transmitted. This leads to significant improvements in terms of energy consumption, and implicitly network longevity.

\begin{figure}[tbp]
\centering
\includegraphics[width=0.999\linewidth,,height=12cm]{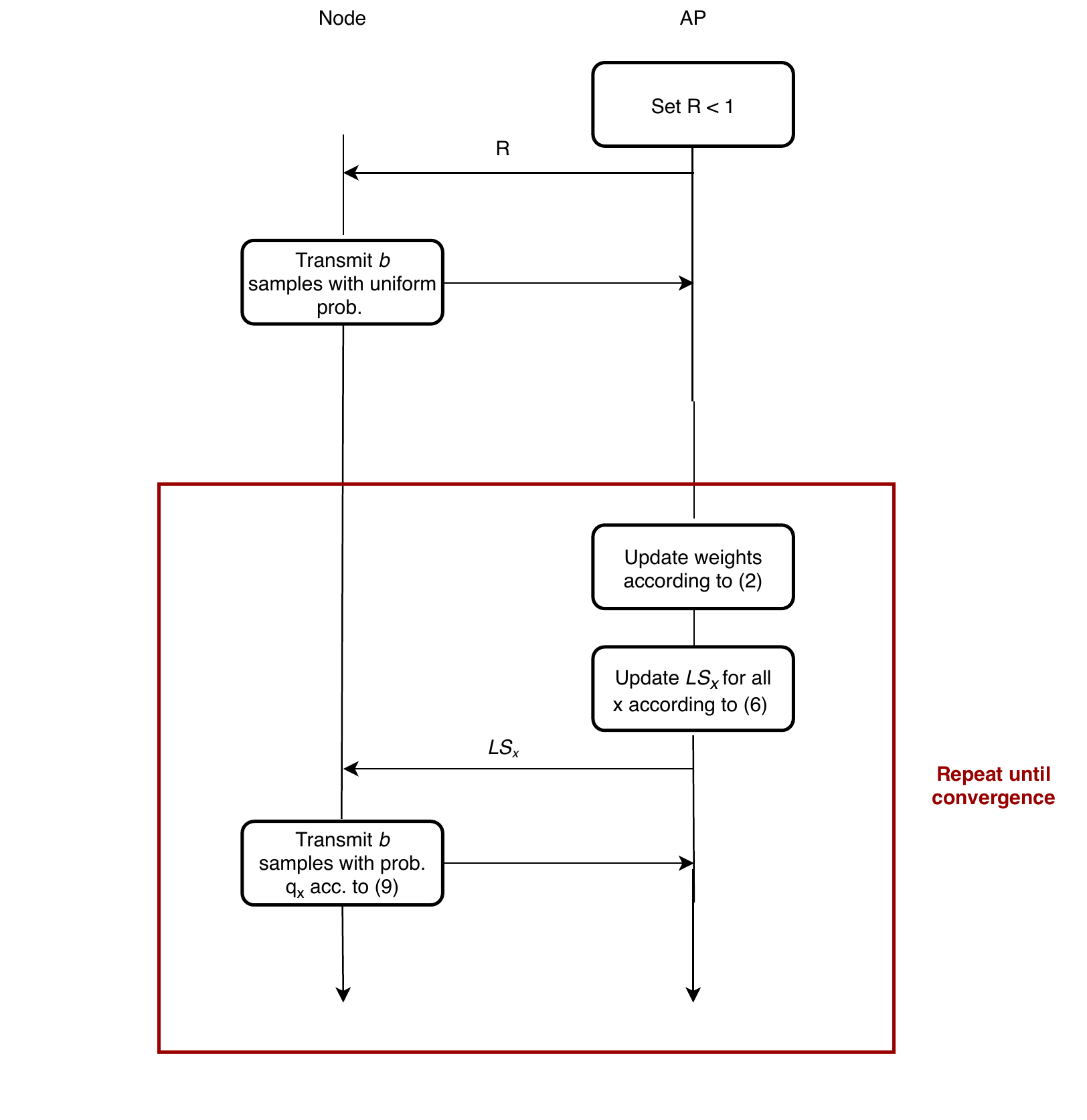} 
\caption{Proposed algorithm.}
\label{fig:algo}
\end{figure}

The only parameters which need to be explicitly defined in the proposed algorithm are the packet rate $R$ and $\mathcal{T}$. The proposed algorithm is given in a flow chart in Fig.~\ref{fig:algo}.
In the spirit of reproducible science, the codes for the algorithm will be made publicly available on \cite{ikoloska}.

\section{Numerical Evaluation}\label{num_eval}
In this section, we evaluate the proposed algorithm. We present the datasets that we use in Section~\ref{data}, and the architecture of the machine learning models in Section~\ref{arch}. The benchmark schemes are described in Section~\ref{bench}, and finally, the results are presented in Section~\ref{poc}.

\subsection{Datasets} \label{data}
We consider three different datasets. In particular, we consider a machine vision dataset, a dataset for leak detection in water distribution networks, and a dataset for  air pollution detection in urban areas.
\subsubsection{Machine Vision}
We consider MNIST \cite{mnist}, a commonly used benchmark dataset for machine vision, in order to train the neural network to recognise digits from $9$ different classes. 
\subsubsection{Leak Detection}
In a water distribution network, water is pumped from the source at one end, through a structure of pipes, to the end users. However, due to external stresors, or accumulated damage, water often leaks in this structure. Such leaks in the pipes are detected by measuring the flow at different points of the distribution network \cite{koo2015towards}, \cite{mutchek2014moving}, \cite{geetha2016internet}. The used datasets are generated by the publicly available simulator EPANET, which is widely used for simulation of water network data~\cite{rossman2000epanet}. 
The architecture of the considered water distribution network is given in \cite{rossman2000epanet}. Thereby, the dataset is comprised of hourly measurements from $9$ links in the water distribution network. In the time series data of flow measurements, leaks are induced according to \cite{fagiani2015novelty}, \cite{boracchi2014exploiting}. 
\subsubsection{Air Pollution Detection}
A large part of the world is experiencing chronic air pollution with severe fine particulate matter concentration ($\text{PM}_{2.5}$ in particular). High levels of $\text{PM}_{2.5}$ are observed and detected via national air pollution monitoring networks. We consider the dataset from the Beijing air pollution monitoring network \cite{zhang2017cautionary}, comprised of hourly measurements from $12$ monitoring stations, in order to train our neural network to detect dangerous levels of $\text{PM}_{2.5}$ in the air.  

\subsection{Architecture of the Machine Learning Models}\label{arch}
We assume that the AP employs uses an artificial neural network as function approximator, $\psi_{\bm{\omega}}(x)$.
For machine vision, the neural network is comprised of $2$ layers with $1200$ hidden units per layer. After each layer, a Dropout layer with probability $0.2$ is included. The logits are then flattened before the output layer.
The neural networks for leak detection and air pollution detection are single layer, $5$ and $32$ hidden unit neural networks, respectively. The learning rates are set to  $0.0005$ and $0.0001$ for leak detection and air pollution detection, respectively. Adam and SGD are the used optimizers for leak detection and air pollution detection, respectively. In addition, in the case of air pollution detection, Dropout with probability $0.2$ is used to regularise the neural network. 
The hyper parameters are summarised in Table~\ref{table1}.

\begin{table*}[]
\centering
\caption{Hyper Parameters}
\label{table1}
\begin{tabular}{llll}
\hline
Hyper Parameters & Leak detection & Air polution & Machine vision           \\ \hline
No. of hidden layers	       & 1      & 1    & 2     \\ \hline
No. of hidden units       & 5      & 32 & 1200        \\ \hline
Learning rate	   & 0.0005		&0.0001	  &0.0001			\\ \hline
Optimizer	   & SGD		& Adam	 &Adam		\\ \hline
Dropout	   & 0		& 0.2   &0.2.			\\ \hline
\end{tabular}
\end{table*}

\subsection{Benchmark Schemes} \label{bench}
We consider \textit{Uniform Filtering} and \textit{Genie-aided Filtering} as benchmark schemes. In order to ensure fairness among the schemes, equal number of packets are transmitted with all schemes.
\subsubsection{Uniform Filtering}
Uniform filtering is straightforward, and relies on uniform selection of packets to be transmitted, i.e., the node transmits a packet with probability
\begin{align}
    q = \frac{1}{M}.
\end{align}
 
\subsubsection{Genie-aided Filtering}
For Genie-aided filtering, we assume the existence of a genie in the network.
The genie knows both the sample and the label $(x,y)$, as well as the class distribution $p(y)=\sum_{x'}p(x',y)$.
Since the genie knows the probability of each class, the genie is able to balance the training dataset and implement a form of importance sampling by choosing to transmit data samples with probability
\begin{align}
    q_{(x,y)} = \frac{\frac{1}{p(y)}}{\sum_{y'} \frac{1}{p(y')}}.
\end{align}
In other words, the genie downsamples the majority class, and upsamples the minority class, and as a result ensures that samples from both classes are transmitted \cite{olson2004data}. This is especially important in cases when the rate is very low. Note that, with uniform sampling often only samples form the majority class are transmitted when the rate is very low.


\subsection{Results} \label{poc}

\begin{figure}[tbp]
\centering
\includegraphics[width=3.4in,height=6.1cm]{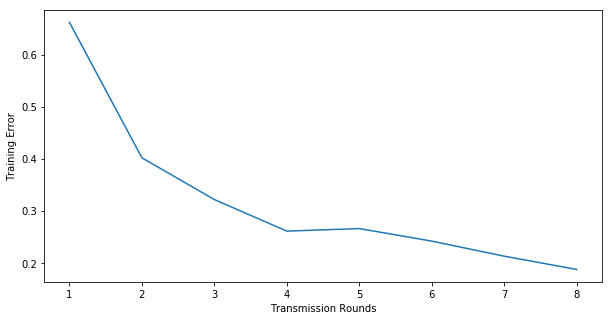}
\caption{Training error as a function of the number of transmission rounds, machine vision. A single run was used for the presented results.}
\label{pof0}
\end{figure}

\begin{figure}[tbp]
\centering
\includegraphics[width=3.4in,height=6.1cm]{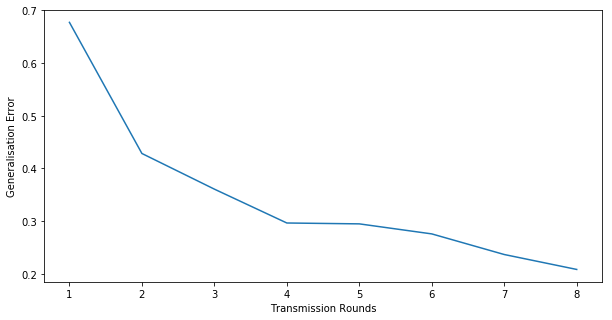}
\caption{Generalisation error as a function of the number of transmission rounds, machine vision. A single run was used for the presented results.}
\label{pof01}
\end{figure}

\begin{figure*}[!htb]
   \begin{minipage}{0.16\textwidth}
     \centering
     \includegraphics[width=.9\linewidth]{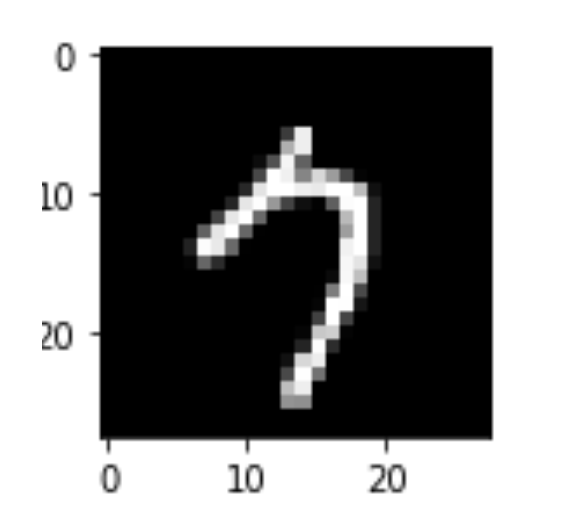}
   \end{minipage}\hfill
   \begin{minipage}{0.16\textwidth}
     \centering
     \includegraphics[scale=.45]{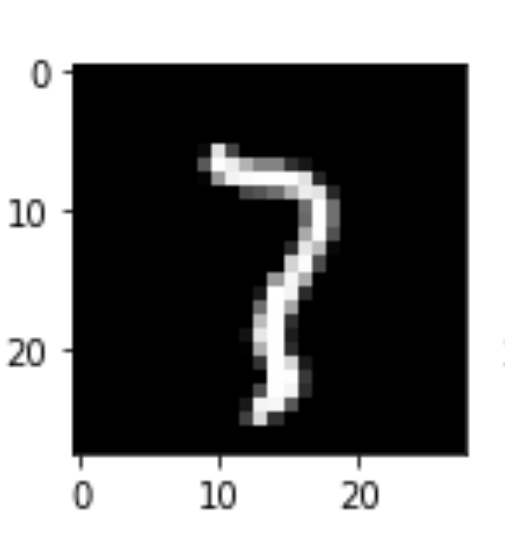}
   \end{minipage}
   \begin{minipage}{0.16\textwidth}
     \centering
     \includegraphics[width=.9\linewidth]{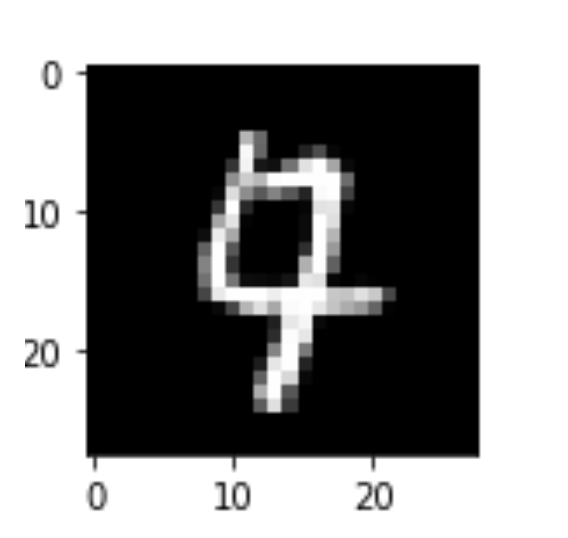}
   \end{minipage}
   \begin{minipage}{0.16\textwidth}
     \centering
     \includegraphics[scale=1.45]{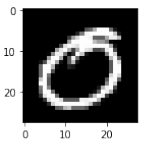}
   \end{minipage}\hfill
   \begin{minipage}{0.16\textwidth}
     \centering
     \includegraphics[width=.9\linewidth]{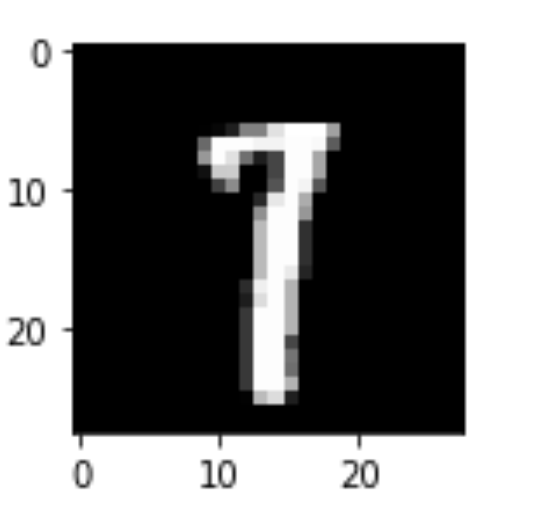}
   \end{minipage}
   \begin{minipage}{0.16\textwidth}
     \centering
     \includegraphics[scale=.45]{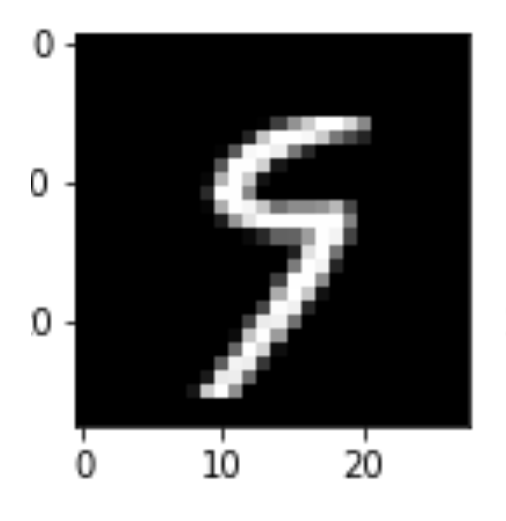}
   \end{minipage}
   \caption{Sample of the transmitted items from the MNIST dataset. The characters appear difficult to classify, which indicate that the filtering scheme works as intended.}
   \label{digits}
\end{figure*}

\begin{figure}[tbp]
\centering
\includegraphics[width=3.4in,height=6.1cm]{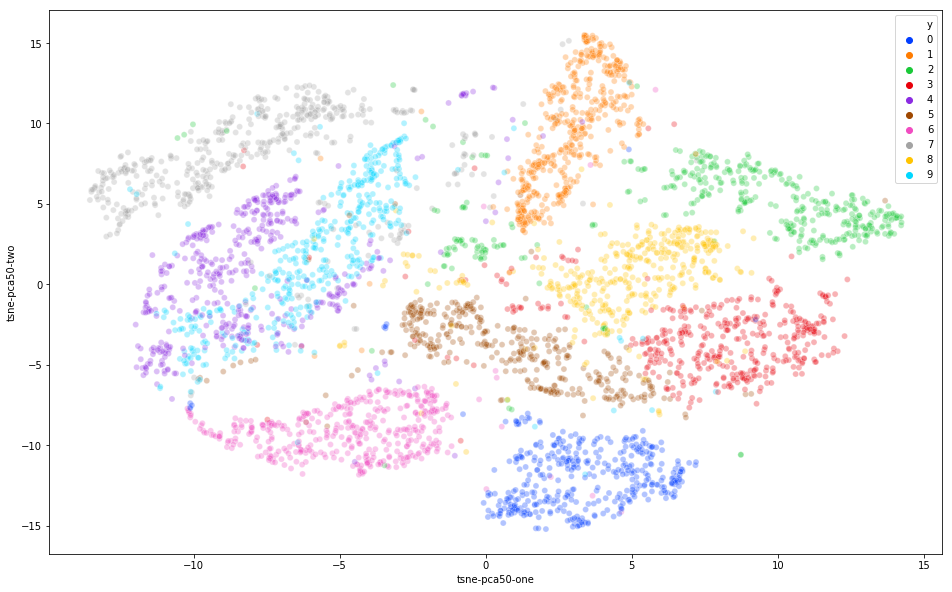}
\caption{t-SNE of all samples, machine vision. A single run was used for the presented results.}
\label{pof1}
\end{figure}

\begin{figure}[tbp]
\centering
\includegraphics[width=3.4in,height=6.1cm]{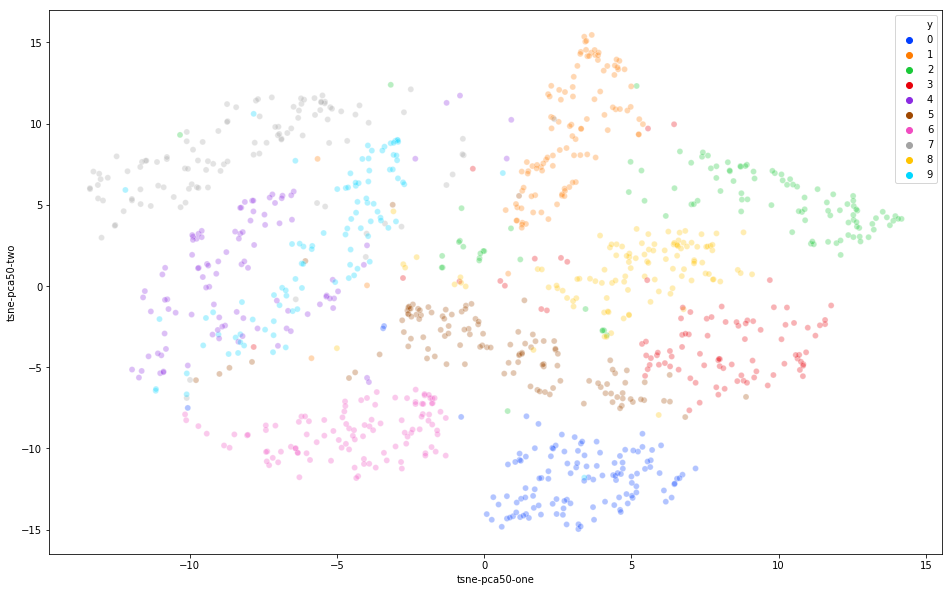}
\caption{t-SNE of samples which have been transmitted, machine vision. A single run was used for the presented results.}
\label{pof11}
\end{figure}

\begin{figure*}[tbp]
\centering
\includegraphics[width=5.4in,height=6.1cm]{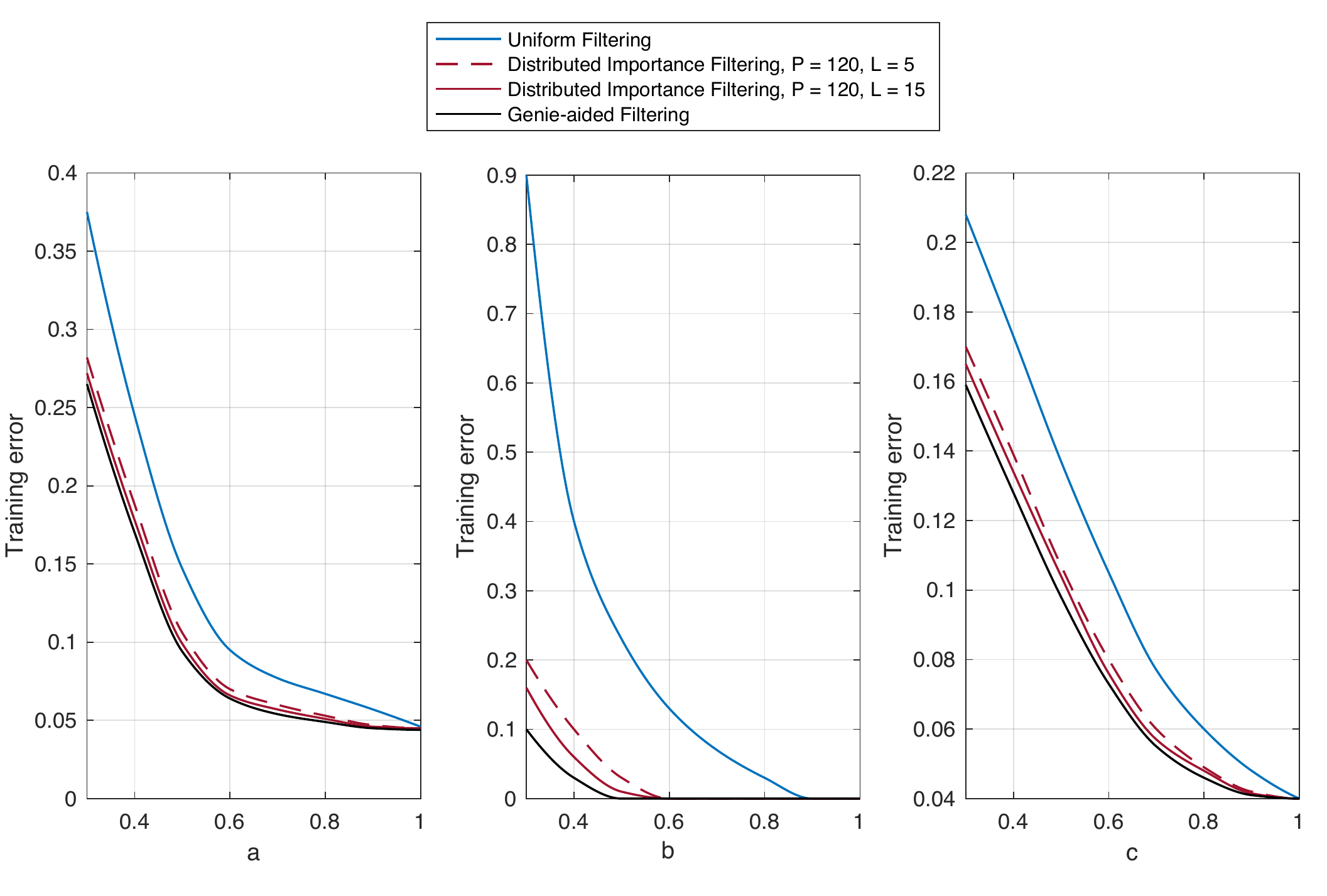}
\caption{Training error (averaged) as a function of the transmission rate $R$ for a) machine vision, b) leak detection, and c) air pollution detection. All presented results are averaged over a) 5, b) 20, and c) 20 independent runs.}
\label{pof_train}
\end{figure*}

\begin{figure*}[tbp]
\centering
\includegraphics[width=5.4in,height=6.1cm]{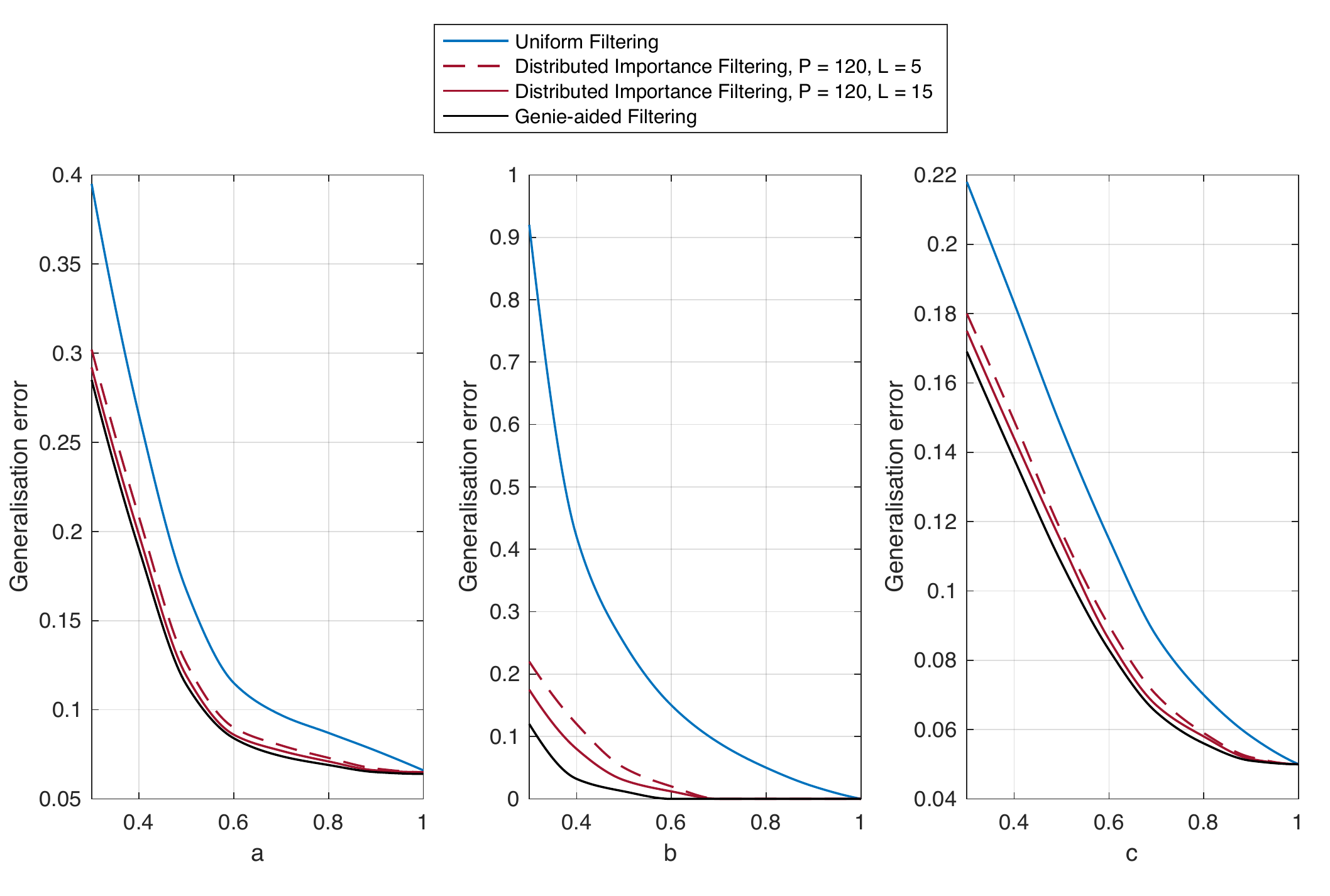}
\caption{Generalisation error (averaged) as a function of the transmission rate $R$ for a) machine vision, b)leak detection, and c) air pollution detection. All presented results are averaged over a) 5, b) 20, and c) 20 independent runs.}
\label{pof_test}
\end{figure*}


We estimate the generalization error by evaluating the loss function of the learned neural networks on a test dataset $\mathcal{D}_{\text{val}}$ drawn from the same distribution as the dataset used for training $\mathcal{D}$, but excluded from the training phase.
We plot the training error as well as the generalisation error for a single run as a function of the transmission rounds for machine vision in Fig.~\ref{pof0} and Fig.~\ref{pof01}, respectively. The rate is set to $R = 0.3$. Initially, in both cases the error drops to $0.45$. However, as the amount of uniform transmissions is annealed and the node starts to greedily transmit only the important samples, the error drastically decreases reaching its minimal value after $10$ transmission rounds. 

Next, we plot a few of the transmitted samples of MNIST in Figure~\ref{digits}. It is interesting to notice that many transmitted samples appear to be digits which are difficult to classify (even for the human eye). For example, the third digit from left to right is a four that can easily be mistaken for a nine, the sixth is a nine that looks like a five etc. To investigate the similarity between the transmitted samples, we plot the t-Distributed Stochastic Neighbor Embedding (t-SNE) of all samples in Fig~\ref{pof1}. The samples which have been transmitted are then extracted and presented in Fig~\ref{pof11}. The reader is referred to \cite{maaten2008visualizing} for details on t-SNE, here it is noted that the main idea is to map high-dimensional data into low-dimensional space by minimizing the divergence between two distributions: a distribution that measures pairwise similarities of the high-dimensional samples and a distribution that measures pairwise similarities of the corresponding low-dimensional points in the embedding. Thereby, even the low-dimensional representation can capture the similarity between the original input samples. In Figs.~\ref{pof1}-\ref{pof11} we map the original 784-dimensional dataset into 2-dimensional space by using fifty principle components generated by the principle component analysis (PCA) reduction algorithm on the original data, and then performing t-SNE. It can be seen that the samples which have not been transmitted form tight, visible clusters and many samples are similar to the samples from their respective class. Conversely, the transmitted samples are very diffused, meaning many samples are ambiguous and are similar to samples belonging to different classes. Using such ambiguous samples for training is beneficial since it refines the decision boundaries of the final model (see for example \cite{zhang2017mixup}). This helps explain the performance gains of distributed importance filtering.

We present the training and the generalisation error of the proposed distributed importance filtering scheme as a function of the packet rate $R$, in Fig~\ref{pof_train}, and Fig~\ref{pof_test}, respectively, for machine vision, leak detection, and air-pollution detection. The number of transmission rounds is set to $10$. As expected, the error decreases when $R$ increases, as more data samples are fed to the model for all schemes. Distributed importance filtering outperforms uniform filtering by a clear margin in all three cases, in spite of using data with different modalities and models with different complexity. Thereby, distributed importance filtering can be seen as agnostic to the input space and the architecture of the model, and can be used in all cases when the gradients can be extracted. In addition, the proposed scheme achieves comparable performance to genie-aided filtering. In fact, both schemes result in the same error with only a small increase of the packet rate for the proposed scheme. The increase of the number of neighbours slightly improves the performance of distributed importance filtering. However, increasing the number of neighbours further may lead to a degradation in performance, as the variance also increases.
 


\section{Discussion and Conclusions}\label{discus}
The proposed scheme in this paper improves the longevity of the network, reduces the
computational complexity for training the model, and eliminates additional data pre-processing
steps. In addition, the algorithm is not exclusively designed to cater to machine vision, or water leakage detection, or air-pollution detection, and as a result it can be used in many real-life scenarios. In fact, the proposed algorithm can be used for any case where the leverage scores can be found according to \eqref{lev_scores}. Straightforward use cases include contamination detection in water distribution networks, gas leak detection, power fluctuations/outage detection in smart grids, blockage detection in oil pipelines etc. 

Extending this work to the more complex case when packet errors in the uplink are taken into
account is rather straightforward, while errors in the downlink packets, which contain feedback from the AP to the nodes, can lead to more involved analysis. In general, a degradation in performance which depends on the
packet error rate is to be expected, both for the proposed scheme, and for the benchmarks. In particular, higher packet rates will be needed to achieve low detection error, as the nodes will need to compensate for the lost pockets. An interesting extension of this work would be to investigate if the nodes can be even more selective when choosing which data samples to transmit, for example by including local cost functions.

\bibliographystyle{IEEEtran}
\bibliography{litdab.bib}
\end{document}